\documentclass[pra,twocolumn,showpacs,nobibnotes,floatfix,superscriptaddress]{revtex4}

\usepackage{latexsym,eucal,amsmath,amssymb,amsfonts,mathbbol,graphicx,color}
\usepackage{dcolumn}% Align table columns on decimal point
\usepackage{bm}% bold math
\usepackage{amsmath}

\newcommand{\ket}[1]{|{#1}\rangle}
\newcommand{\bra}[1]{\langle {#1}|}

\begin{document}

%\preprint{}
%\draft

\title{Parameters of Pseudo-Random Quantum Circuits}

\author{Yaakov S. Weinstein}
\thanks{Electronic addess: {\tt weinstein@mitre.org}}
\affiliation{Quantum Information Science Group, {\sc Mitre},
260 Industrial Way West, Eatontown, NJ 07224, USA}

\author{Winton G. Brown}
\thanks{Electronic addess: {\tt Winton.G.Brown@Dartmouth.EDU}}
\affiliation{Department of Physics and Astronomy, Dartmouth
College, Hanover, NH 03755, USA}

\author{Lorenza Viola}
\thanks{Electronic addess: {\tt Lorenza.Viola@Dartmouth.EDU}}
\affiliation{Department of Physics and Astronomy, Dartmouth
College, Hanover, NH 03755, USA}

\date{\today}

\begin{abstract}
Pseudorandom circuits generate quantum states and unitary operators
which are approximately distributed according to the unitarily
invariant Haar measure.  We explore how several design parameters
affect the efficiency of pseudo-random circuits, with the goal of
identifying relevant trade-offs and optimizing convergence.  The
parameters we explore include the choice of single- and two-qubit
gates, the topology of the underlying physical qubit architecture, the
probabilistic application of two-qubit gates, as well as circuit size,
initialization, and the effect of control constraints. Building on the 
equivalence between pseudo-random circuits and approximate  
$t$-designs, a Markov matrix approach is employed to analyze
asymptotic convergence properties of pseudo-random second-order
moments to a $2$-design.  Quantitative results on the convergence rate
as a function of the circuit size are presented for qubit topologies
with a sufficient degree of symmetry.  Our results may be
theoretically and practically useful to optimize the efficiency of
random state and operator generation.
\end{abstract}

 % Quantum computation

\maketitle

\section{Introduction}

Random pure states play a prominent role in quantum
information processing.  Random states are defined with respect to
the unitarily invariant (so-called Fubini-Study) measure on the
space of unit vectors in the Hilbert space of the system
\cite{KarolBook}.  Not only do random states possess the
remarkable feature of saturating the classical communication capacity
of a noisy quantum channel \cite{Seth2}, they also are an enabling
resource for protocols including superdense coding of quantum
states \cite{Aram}, approximate quantum encryption,
and quantum data hiding \cite{Hayden,Smith}.
Random quantum states can also be used for unbiased sampling, and
the characterization of both `typical' bipartite and multipartite
entanglement in random states has long been the focus of extensive
investigation \cite{Page,Foong}, recent results including estimates
of moments of the subsystem purity distribution
\cite{Scott,Giraud07}, relationships to state localization
properties \cite{WintonJPA},
and exact expressions for the probability distribution of
entanglement measures such as
$G$-concurrence \cite{Cappellini} and purity \cite{Facchi}.

How does one generate random quantum states? One way is to apply
a random unitary transformation to initial computational basis states.
Similar to random states, random unitary operators are drawn uniformly
from the unique unitarily invariant measure -- the Haar measure
on the unitary group U($N$), where $N$ denotes the Hilbert space
dimension, $N=2^n$ for a multipartite $n$-qubit system \cite{Mehta}.
Random unitaries themselves are useful for a number of
quantum protocols ranging from remote state preparation \cite{Bennet} to
efficient error characterization \cite{RM,Levi} and selective process
tomography \cite{Paz}.  Unfortunately, implementing random unitaries as a
sequence of one- and two-qubit gates on a quantum computer
is inefficient: the required number of quantum gates grows quadratically
with $N$, that is, exponentially with the number of qubits, $n$.
Therefore, researchers are left with the challenge of constructing
suitable pseudo-random (PR) substitutes.

A promising approach to efficiently implement an ensemble of unitaries
so that the resulting moments approximate those induced by the Haar
measure is provided by {\em PR circuits}, introduced in Ref.
\onlinecite{RM}.  These circuits consist of an iterated set of single-
and two-qubit gates where certain specifications are chosen at random.
As the set of gates is iterated (using different single-qubit gates
for each qubit and at each time step), the statistical properties of
the implemented unitary operators and the resulting output states
compare more and more favorably to the properties of random unitaries
and states.  Additional studies of PR circuits have focused on
analyzing both analytically and numerically the convergence to the
desired distribution \cite{ELL,YSW1,Braun08}, identifying optimal
two-qubit gates, as well as elucidating some aspects related to the
influence of qubit topology \cite{Znidaric}, and quantifying the
ability to efficiently generate states with generic entanglement
\cite{ODP}.  PR algorithms have also been formulated for cluster-state
quantum computation (QC) in Ref.  \onlinecite{BWV}, allowing an
optimal single-gate distribution to be identified.  A yet different
realization via local measurements on weighted graph states has been
proposed in \cite{PCP}.

Our goal in this paper is to quantitatively investigate a number of
parameters that affect the convergence of PR circuits to the Haar
distribution.  After providing the necessary background on the
mathematical framework employed to characterize PR behavior in
Sec. II, we proceed in Sec. III to assess the influence of three main
design parameters in a PR circuit on a fixed number of qubits: the
choice of single- and two-qubit gate distributions; the influence of
different qubit topologies; and the effect of probabilistic versus
deterministic two-qubit gate application.  Not surprisingly, these
parameters are intertwined, causing the optimal choice for any one
parameter to depend on one or several other parameters in complex
ways.  The influence of circuit size is addressed in Sec. IV.
Remarkably, explicit scaling predictions turn out to be possible based
on simple analytical expressions which are consistent with existing
numerical evidence and analytical results. Secs. V and VI are devoted
to analyze the effect of limited (non-selective) control and of
different initial states, respectively. In Sec. VII we make some final
remarks and conclude. Additional considerations on optimizing
cluster-state PR circuits are included in Appendix A, whereas Appendix
B includes a proposed convergence improvement in the specific yet
important case of a PR circuit implementing an approximate Clifford
twirl.

%%%%%%%%%%%%%%%%%%%%%%%%%%%%%%%%%%%%%%%%%%%%%%%%%%%%%%%%%%%%%%%%%%%%%%%%%%%%

\section{Quantifying pseudo-randomness}

We begin by discussing possible ways of quantifying the distance
between the ensemble generated by PR circuits and the Haar-distributed
ensemble. It is important to realize that any probability distribution
over $n$-qubit quantum states or unitary transformations requires a
number of parameters exponentially growing with $n$ to
specify. Therefore, it is impractical to gauge how well an ensemble of
quantum states or unitary transformations resembles the uniform Haar
ensemble based on a full characterization of the distribution.

\subsection{$t$-designs}

For several tasks which utilize random states and unitary
transformations, the details of the full distribution are not
relevant, in the sense that only statistical moments up to a {\em
finite} order of the ensemble from which the random states or
unitaries are drawn need to coincide with Haar-induced moments.  That
is, it suffices that the relevant probability distribution be
indistinguishable from the uniform Haar distribution as long as a
finite number of moments is given.  This is captured by the concept of
a {\em quantum $t$-design} \cite{EmersonDesign}.  Formally, an
ensemble of states, $[p(\alpha)d\alpha$,$\ket{\psi({\alpha})}]$, (or,
respectively, unitary transformations, $[p(\alpha)d\alpha, U(\alpha)]$)
is a state (unitary) $t$-design if for any polynomial $f$ of order
($t$,$t$) of the state-vector components (or matrix elements),
$$\int p(\alpha) f(\ket{\psi(\alpha)} ) d\alpha
= \int f(\ket{\psi(\alpha)} ) d\alpha , $$
or,
$$\int p(\alpha) f(U(\alpha)) d\alpha =
\int f(U(\alpha)) d\alpha ,$$
\noindent
where the integrals are taken with respect to the invariant
(Fubini-Study or Haar) measure, respectively.  A ($t$,$t$) polynomial
refers to a linear combination of terms consisting of products of up
to $t$ variables and their $t$ complex conjugates.

In order to assess how well a PR circuit approximates a $t$-design, an
appropriate norm must be defined on the space of ($t$,$t$) polynomials
describing moments of the pseudo random circuit. We note that the
concept of an $\epsilon$-approximate $t$-design for quantum states was
introduced in \cite{EmersonDesign} as an ensemble of states
$[p(\alpha)d\alpha$, $\ket{\psi({\alpha})}]$ satisfying
\begin{eqnarray*}
(1-\epsilon) \int
f(\ket{\psi(\alpha)}) d\alpha &\leq &
\int \hspace*{-1mm}
p(\alpha) f(\ket{\psi(\alpha)}) d\alpha \\
& \leq & (1+\epsilon) \int
f(\ket{\psi(\alpha)}) d\alpha,
\end{eqnarray*}
\noindent
for all ($t$,$t$) polynomials $f$. This notion induces a norm on the
space of all ($t$,$t$) polynomials (specifically the $l_\infty$-norm),
which can quantify the distance of a distribution approximating an
exact $t$-design. In our study, we shall focus on asymptotic convergence \textit{rates}
of arbitrary (2,2) polynomials to their expected value under the Haar measure
as a function of PR circuit depth.  Note that Ref.~\cite{Harrow08} has
previously shown that a large class of PR
circuits are efficient approximate unitary $2$-designs, with respect to
the diamond norm.

\subsection{Markov Chain analysis}

The case $t=2$ is especially relevant, since it includes the majority
of known protocols which utilize random states. Notable examples
include unbiased noise estimation \cite{Levi,Paz}, and the generation
of states with typical entanglement \cite{ODP}. It is thus important
to determine the behavior of $(2,2)$ polynomials (second moments) of
the state components under the action of PR circuits.  As shown in
\cite{ODP}, the evolution of the second moments can be mapped to an
appropriately defined classical Markov chain.  Thus, convergence
properties of the PR circuit to a $2$-design may be directly
established by exploiting properties of the corresponding Markov
chain.

To obtain the mapping, the density operator describing the pure
quantum state being evolved by the PR circuit is written in the
Pauli basis: $$\rho =
|\psi\rangle\langle\psi| = \sum_\nu c_\nu P_\nu,$$
where $P_\nu =
\sigma_1^{\nu_1} \otimes \ldots \otimes \sigma_n^{\nu_n}$ is a
tensor-product string
of single-qubit identity and Pauli operators, specified by the
collective index $\nu \in {\mathcal I}=\{0,x,y,z \}^n$.  Let
PR$(\ell)$ be the family of PR circuits of depth $\ell$. We shall be
interested in the evolution of the second order moments, $\{{\mathbb
E}_{\text{PR}(\ell)}(c_{\nu,\ell}c_{\mu,\ell})\}$, as a function of the
depth, $\ell$, of the PR circuit.  It will be shown by
construction that the moments $\{{\mathbb
E}_{\text{PR}(\ell)}(c_{\nu,\ell}^2)\}$ follow a
discrete Markov chain on ${\mathcal I}$.
That is, the evolution of ${\mathbb
E}_{\text{PR}(\ell)}(c_{\nu,\ell}^2)$ satisfies
$$ %\begin{equation}
{\mathbb E}_{\text{PR}(\ell+1)}(c_{\nu,\ell+1}^2)=
\sum_{\nu \in {\cal I}} M_{\mu \nu}
{\mathbb E}_{\text{PR}(\ell)}(c_{\nu,\ell}^2)=
\sum_{\nu \in {\cal I}} M_{\mu \nu}^\ell c_{\nu,0}^2,
$$
where $M=\{M_{\mu \nu}\}$ is a Markov matrix. For the random circuits
examined here, the remaining second moments $\{{\mathbb
E}_{\text{PR}(\ell)}(c_{\nu,\ell}c_{\mu,\ell})\}$, with $\nu \neq \mu$,
vanish for $\ell >0$.

The class of PR circuits we are interested in have the following
structure: In a
single time step, local (single-qubit)
gates are applied to each qubit in parallel,
followed by commuting two-qubit gates belonging to the 
Clifford group between a specified set of
neighboring qubits. Typically, the two-qubit gates will be 
conditional phase gates CZ$=\ket{0}_1\bra{0}_1 \otimes \openone_2 +
\ket{1}_1\bra{1}_1\otimes \sigma_z^2$.
Under a single-qubit gate, each non-trivial Pauli
operator transforms as
$$\sigma_a\mapsto R(\sigma_a)=\sum_{a,b} x_{ab}\sigma_b,\;\;\; a,b \in
\{x,y,z\},$$ where $R=\{x_{ab}\}\in$ SO(3) depends on the applied
rotation. The corresponding transformation matrix $\overline{R}$
acting on the space of local Pauli operators (including the identity,
$\openone$) is obtained by averaging the squared coefficients over the
parameters specifying the desired distribution of rotations. We
examine only those local gate distributions that satisfy
$${\mathbb E}(x_{ca}x_{cb})=0,\;\;\; \forall a,b,c.$$
\noindent
This condition ensures that $\overline{R}=\openone
\oplus {\mathbb E}(x_{ab}^2)$ is indeed a Markov matrix.  Assuming that each
local gate is {\em selected independently} from the same distribution (see Sec.
V for a different setting), the transformation resulting from
simultaneous single-qubit rotations within a PR iteration is the
$n$-fold tensor product of these single-qubit transformations, $L =
\overline{R}^{\otimes n}.$ Since each two-qubit gate is a member of the Clifford
group, it simply acts (up to irrelevant phases) as a permutation on the columns
of $L$. Thus, the full $4^n$-dimensional transformation is a Markov matrix
provided that $\overline{R}$ is.

Once the mapping is established, basic properties of Markov chains may
be used to analyze the evolution of the second moments of the PR
circuit. Specifically, if the Markov chain is {\em ergodic}, then the
corresponding PR circuit converges to a $2$-design.  A Markov chain is
ergodic if it is {\em irreducible} and {\em aperiodic}.  That is,
every state must be reached from every other state and the recurrence
times to any state must not be multiples of some period $k > 1$. A
consequence of ergodicity is that for any initial distribution $v_o$
over ${\mathcal I}$, there exists a unique asymptotic distribution,
$v_\infty$, towards which the chain evolves.  To illustrate this, we
may expand the initial state in terms of the (linearly independent)
eigenvectors $\{e_n\}$ of the Markov matrix, $v_\tau=M^\tau v_o =
\sum_n \lambda_n^\tau c_n e_n$.  For an ergodic chain, there is only
one eigenvalue equal to $1$, whereas all other eigenvalues have
magnitude less than 1. Thus, the contribution to $v_\tau$ from each
eigenvector (other than the ergodic eigenvector that is equal to 1)
decays to zero exponentially at a rate governed by the corresponding
eigenvalue. For times $\tau \gg \log({\lambda_k}/{\lambda_1})$, the
contribution from the sub-dominant eigenvalue, $\lambda_1$, will be
the dominant non-ergodic contribution, leaving
\begin{equation}
\label{ExpGamma}
||v_{\tau}-v_\infty||_\infty \sim e^{-\Gamma \tau},
\end{equation}
where $\Gamma = - \ln(\lambda_1)$, and $\Delta=1-\lambda_1$ is the
spectral gap of the Markov chain.  Since any (2,2) polynomial may be
expressed in terms of the moments $\{{\mathbb
E}_{\text{PR}(\ell)}(c_{\nu,\ell}c_{\mu,\ell})\}$ it follows that at
sufficiently long times, $\tau$, the difference between the expected
value of any $(2,2)$ polynomial of the state vector components over
the $t$-design distribution and the Haar-induced expected value obeys
\begin{equation}
|f_{2{\text{-design}}}
(\tau)-f_{\text{Haar}}(\tau)| \sim e^{-\Gamma \tau }.
\end{equation}
As we shall see, certain entanglement measures which serve as useful
test functions for PR circuits are expressible as $(2,2)$ polynomials
and will thus converge exponentially to their Haar-expected value at a
rate of $\Gamma$.

\subsubsection{Markov chain reduction}

A Markov chain may be reduced by identifying a partition of the state
space, $\mathcal{I}$, into subsets, $\mathcal{J}_u\subset\mathcal{I}$,
such that the coarse-grained probability distribution obtained by
summing over each subset follows a reduced Markov chain $M'$. A
necessary and sufficient condition for reducing the chain is that the
sum of the transition probabilities from a member of a subset, $\mu
\in \mathcal{J}_u$, to
all members of any subset, $\nu\in \mathcal{J}_v$, is the same for
each member $\mu$,
that is, $\sum_{\mu,\nu \in \mathcal{J}_v} M_{\mu\nu}
= M'_{uv}$ must be the same for
all ${u\in \mathcal{J}_u}$ and all subsets $\mathcal{J}_v$.
%%LV: Winton, I presume that's what you meant?
It is important to note that each eigenvalue of $M'$ is an eigenvalue
of $M$.

We employ a reduced representation by averaging over local $X$ and $Y$
Pauli operators.  Since CZ gates are invariant under rotations along
the $z$-axis, we may restrict our choice of single-qubit gate
distributions to those which initially randomize states in the $x$-$y$
plane. Now let $P$ be a Pauli string containing at least one $X_i$ or
$Y_i$, and let $P'$ be any string obtained from $P$ by permuting $X_i$
with $Y_i$. Since $M$ randomizes $X_i$ and $Y_i$, $M(P-P') = 0$. This
defines the kernel of $M$, which may be removed by defining new
variables $\Xi_i^\pm = X_i \pm Y_i$.  Chain states including
{$\Xi_i^-$} may be discarded, whereas transitions within ${\cal I}'=
\{0,z,\xi\}^n$ are described by $M'$.  There is now only one parameter
left to characterize the local gate distribution.  Let $c \in [0,1]$
parameterize the extent to which the $z$-axis is left invariant. We
shall refer to $c$ as the {\em local gate parameter} henceforth.  The
single-qubit gate contribution to $M'$ is then fully described by:
\begin{eqnarray*}
\overline{R}(c)= \left(
  \begin{array}{cccc}
    1 & 0 & 0  \\
    0 & c & \frac{1-c}{2} \\
    0 & 1-c & \frac{1+c}{2} \\
  \end{array}
\right).
\end{eqnarray*}
Haar-distributed rotations in SU$(2)$ correspond to $c = 1/3$.
So-called {\em HZ gates},
single-qubit Hadamard gates followed by a random rotation about the
$z$-axis, correspond to $c = 0$.  The latter were identified
in \cite{BWV} as optimal single-qubit gates when using two-qubit
CZ gates.

In some instances, it is possible to further reduce the Markov chain
by taking into account qubit-permutation symmetries in the
construction of PR circuits (see Sec. IV). The corresponding Markov
chain will admit a reduced chain by forming equivalence classes of
Pauli strings under symmetry operations. In the case of {\em full
permutation symmetry}, equivalence classes may be labeled by the
number of $X$'s and $Z$'s in each Pauli string.  Accordingly, the size
of the reduced state space grows only quadratically in the number of
qubits $n$, allowing scaling behavior of Markov matrix properties with
circuit size to be determined numerically.  Note that while the
reduced representation enabled by the initial randomization of $x$ and
$y$ contains all of the non-zero eigenvalues of the full Markov chain,
reduced representations induced by qubit permutations discard
eigenvalues associated with asymmetry with respect to the
permutations. Thus, the behavior of test functions which do not
possess the symmetry cannot be predicted if the initial state does not
also share the symmetry which permits the reduction.

%%%%%%%%%%%%%%%%%%%%%%%%%%%%%%%%%%%%%%%%%%%%%%%%%%%%%%%%%%%%%%%%%%%%%%%%%%%%

\subsection{Entanglement measures}

As remarked, entanglement properties of random pure states have been
extensively studied. In particular, bipartite-entanglement across a
partition is known to be nearly maximal for random states.  As
observed in \cite{ODP}, linear entropy, a measure of bipartite
entanglement, may be expressed as a $(2,2)$ polynomial in the state
components, allowing the rate at which bipartite entanglement
approaches its expected Haar-value to be determined by Markov
analysis.  Both features -- that the asymptotic value of the
entanglement is nearly maximal, and that the simplest measure of the
entanglement is a $(2,2)$ polynomial in the state vector components --
are not limited to bipartite entanglement but also apply to a more
general class of entanglement properties that are captured by the {\em
Generalized Entanglement} (GE) approach \cite{Barnum}.

The basic idea of GE is to abstract the notion of
entanglement from a preferred subsystem decomposition and instead tie
it to a distinguished set of observables to which one has access
(in an appropriately defined sense).
Intuitively speaking, a pure quantum state $\ket{\psi}$ is
defined as `generalized unentangled' relative to a distinguished
set of observables, $h$, depending on whether or not the `reduced state'
with respect to those observables is pure (extremal), and generalized
entangled otherwise. That is, if one can
determine which pure state a system is in by only making reference
to the expectation values of observables in $h$,
then the reduced state is pure.  Under appropriate mathematical
assumptions on $h$ \cite{GE2}, the simplest way to
quantify GE is by taking the square length of the projection
of the full pure state $|\psi\rangle\bra{\psi}$ onto $h$, yielding
the so-called $h$-purity.  Formally, the latter
is defined as
$P_h(\ket{\psi})=\kappa \sum_i \bra{\psi} A_i \ket{\psi}^2,$
where $\kappa >0$ is a normalization constant and \{$A_i$\} is
an orthonormal (with respect to the trace norm) basis of
$h$. GE relative to $h$ may be
naturally quantified as GE$_h=1-P_h$. A simple representative of this
class of state functionals in the conventional multi-partite setting
is the global entanglement measure of Meyer and Wallach
\cite{Meyer,Barnum,Bren2}:
\begin{equation}
\label{Q}
Q(\ket{\psi}) = 2-\frac{2}{n}\sum^n_{j=1}{\text Tr}[\rho_j^2],
\end{equation}
where $\rho_j$ is the reduced density matrix of qubit $j$.
The expectation of $Q$ over random pure states is
\cite{Scott,WintonJPA}:
\begin{equation}
\langle Q_R\rangle = \frac{N-2}{N+1}.
\label{Qhaar}
\end{equation}
The difference between the average entanglement of PR algorithm output
states and random state entanglement,
$|\langle Q\rangle -\langle Q_R\rangle|$,
provides a distance indicator between PR and random states
whose asymptotic behavior should be
predicted by the Markov matrix analysis. The average
Meyer-Wallach entanglement has been examined in previous studies of
PR circuits \cite{RM}, further motivating its consideration in the
present context.

\subsection{State-vector element distribution}

The other method we use to assess the quality of our PR circuits is
based on the distribution of state vector probability in the
computational basis.  Let $c^l_k$ denote the $k$th component of the
$l$th state randomly chosen from the set of all pure states, with
$\eta = |c^l_k|^2$ being the corresponding probability.  The
distribution of $\eta$ for random states is given by the Porter-Thomas
(PT) distribution \cite{KarolBook,Zyc2}:
$$ \tilde{P}(\eta) = (N-1)(1-\eta)^{N-2}, $$
where $N$ is the Hilbert space dimension. In the limit
$N \rightarrow \infty$, and upon rescaling to unit mean, the distribution becomes
$$ P_{PT}(y) = e^{-y}, \;\;\;y = N\eta. $$ As our second randomness
indicator, we shall examine the $l_2$-distance (simply denoted by $||
\cdot ||_2 \equiv |\cdot |$ henceforth) between the distribution of
output probabilities at each PR iteration and the random
(Porter-Thomas) distribution.  Note that because this test function
depends on higher order moments of the PR-distribution, it is not {\em
a priori} described by the above-described Markov chain analysis and
thus can yield additional insight.

%%%%%%%%%%%%%%%%%%%%%%%%%%%%%%%%%%%%%%%%%%%%%%%%%%%%%%%%%%%%%%%%%%%%%%%%%%

\section{Optimizing pseudo-random circuits}

Any universal set of gates used in a PR circuit will eventually
generate Haar-distributed states and operators \cite{ELL}. Our goal
here is to explore how various design parameters affect PR circuits,
in an effort to boost efficiency by optimizing time and qubit
resources.  By focusing on the network model of QC, we first
characterize the optimal choice of single- and two-qubit gates for an
open-chain topology as considered in the `standard' PR architecture of
Ref. \onlinecite{RM}.  The effect of different qubit topologies and
probabilistic gate application is addressed next, whereas, for clarity,
a similar analysis for cluster-state QC is sketched in Appendix A.

\subsection{Optimal gates for standard circuits}
PR circuits were introduced in the context of liquid-state nuclear
magnetic resonance quantum information processing.  In that context it
was natural to look at an open chain of qubits and to use two-qubit
Ising (ZZ) gates between nearest neighbor (NN) qubits,
$$ {\text
{ZZ}}=\exp\Big[-i\frac{\pi}{4}\sigma_z^j\sigma_z^{j+1}\Big],$$ as the
relevant non-local resource.  Subsequently, other two-qubit gates have
been suggested for use within PR algorithms \cite{ODP}.  A
comprehensive study was recently undertaken \cite{Znidaric} to
determine the two-qubit gate for which a PR circuit with random
single-qubit gates will converge most quickly to the Haar expected
value of bipartite entanglement as quantified by linear
entropy. Recall that this is essentially equivalent to determining
convergence rates of second order moments of the PR circuit.  The
protocol studied calls for a single two-qubit gate to be applied per
time interval to pairs of qubits which are selected at random from all
allowable neighboring pairs according to either a closed or open chain
\cite{RemarkXY}. It was determined that the so-called {\em XY gate} is
the two-qubit gate which leads to the most efficient PR circuit of
this type. The XY gate between qubits $j$ and $k$ is given by
$${\text {XY}}=
\exp\Big[-i\frac{\pi}{4}(\sigma_x^j\sigma_x^{k}+\sigma_y^j\sigma_y^{k})
\Big]. $$ However, the use of only a single two-qubit gate per time
interval introduces an unnecessary inefficiency since qubits not
involved in the gate do nothing for that time interval and, in any
experimental implementation, must be protected from error.

\begin{figure}[t]
\includegraphics[width=8cm]{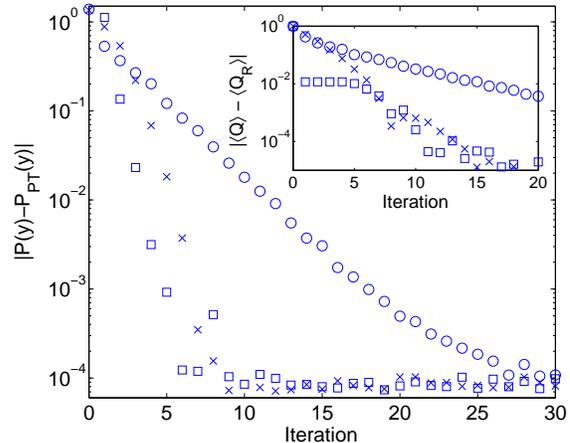}
\caption{(Color online)  $l_2$-distance between PR distribution of
squared moduli of components in the computational basis and the
Porter-Thomas distribution (Inset: Distance of average global
entanglement from random value) for
CZ ($\bigcirc$) and XY ($\times$) two-qubit gates using single-qubit
random rotations, and CZ gates using single-qubit HZ gates ($\square$)
(8 qubits, 100 implementations, all computational basis states).
When using random single-qubit gates, both indicators come within
$10^{-4}$ much more quickly for
XY two-qubit gates than for CZ gates.  Modifying the CZ-based
algorithm to use single-qubit HZ gates boosts the convergence rate
so much that it even outperforms the
XY-based algorithm with random single-qubit rotations.}
\label{ZZchain}
\end{figure}

We wish to examine to what extent the number of applied gates can be
traded for the number of time steps by applying gates in parallel.
One cannot apply all NN XY gates during one time interval since they
do not commute. Nevertheless, a more efficient PR circuit using XY
gates can be obtained if, at each time interval, half the NN XY gates
are applied. Fig. \ref{ZZchain} compares the state vector element
distribution distance and average entanglement distance as a function
of iteration for a PR algorithm which uses conditional phase (CZ)
gates (which are ZZ gates to within single-qubit $z$-rotations)
between all NN qubits, and a PR algorithm which uses XY gates between
half the NN pairs at each iteration (say, 1-2, 3-4, etc.~at odd
iterations and 2-3, 4-5, etc.~at even iterations). In both cases, the
single-qubit rotations are drawn randomly from SU$(2)$.  It takes
significantly more time for the CZ-based PR algorithm to be within
$10^{-4}$ of typical entanglement values and PT distribution
than the XY-based PR algorithm, despite the fact that all NN CZ gates
are applied at every iteration.

In all work mentioned above, it was assumed that single-qubit
rotations should be random with respect to the Haar measure on
SU$(2)$.  Based on studies of cluster-state PR algorithms, however, we
have recently shown \cite{BWV} that when using two-qubit CZ gates a
{\em restricted set of single-qubit rotations}, HZ gates, allows for
faster convergence with respect to test functions which are
second-order polynomials. This is confirmed through the Markov
analysis reported in Fig. \ref{Mtop} which shows that, for
sufficiently large $n$, the optimal single-qubit gates to use in
conjunction with the CZ gate is indeed at $c = 0$, the HZ gate.  Using
these restricted random gates we see in Fig. \ref{ZZchain} that the
CZ-based PR circuit converges much more quickly than the CZ-based
circuit with random SU$(2)$ single-qubit gates.

What single-qubit gate distribution is optimal for use with XY gates?
From the
inset of Fig. \ref{Mtop}, we see that the answer depends
on the qubit topology. For qubits on an open chain, as in standard
PR circuits, the optimal single-qubit gate is at $c \simeq .5$.
For a closed-chain topology,  SU(2) random
single-qubit rotations appear to be optimal.

\subsection{Effect of different qubit topologies}

\subsubsection{Modeling strategies and preliminaries}

As mentioned, a standard PR circuit \cite{RM} applies two-qubit gates
between NN qubits on an open chain. Recent studies of PR algorithms
have implemented two-qubit gates between NN qubits on a closed chain
({\em i.e.}, subject to periodic boundary conditions) \cite{Znidaric},
or between randomly chosen qubit pairs \cite{ODP,MSB}. This alters the
convergence rate of the algorithm by effectively changing the qubit
topology.  In this section, we explore a number of representative
topologies and the convergence rate of the implemented circuits as a
function of different single-qubit gate distributions.

Specifically, we investigate the following alternate topologies which
employ two-qubit CZ gates: (i) NN qubits on a closed chain; (ii) a
star formation where each qubit is connected to a central one; (iii)
an all-to-all (AA) topology where CZ gates are performed between every
possible pairing of qubits (note that all such gates commute and can
thus be performed simultaneously).  Our strategy is to rely on Markov
chain analysis to explore how different single-qubit gate
distributions affect the convergence rate for different topologies,
thereby identifying the optimal single-qubit gate for each topology.
For PR circuits employing XY gates, open- and closed-chain NN
topologies are contrasted, as a function of $c$.

The circuits studied in this subsection aim at {\em maximizing
parallelism}, by applying as many CZ gates as possible per
iteration. In this case, the closed-chain topology is {\em not}
ergodic over the entire state space. This can be seen by noticing that
each qubit undergoes exactly two CZ gates per iteration (one with each
NN). Under the action of pairs of CZ gates, the number of non-identity
single Pauli operators in the Pauli string describing the state will
remain even or odd.  Within each of these fixed-parity subspaces, the
closed-chain topology is ergodic and the PR circuit will induce
exponential convergence to the corresponding ergodic state, with a
rate calculated from the gap of the Markov matrix.  The asymptotic
entanglement for an $n$-qubit closed chain is also affected by the
presence of the two distinct subspaces. Specifically, the total
probability that the state of the Markov chain is in each subspace
remains constant and is determined by the projection of the initial
state into each subspace. Under the action of the Markov chain,
whatever total probability is found in each subspace is uniformly
distributed within each subspace, allowing for the asymptotic
entanglement to be calculated. For initial computational basis states,
the resulting closed-chain (cc) global entanglement is given by:
\begin{equation}
Q_{\text{cc}} = 1-3\frac{\sum_{k \;\rm{odd}}^{n} {n \choose k}}{\sum_{k
\;\rm{odd}}^n 3^k {n \choose k}}.
\label{CCEnt}
\end{equation}
For $n = 8$, $Q_{\text{cc}} = .988235$, whereas the expected global
entanglement value for the Haar distribution (from Eq. (\ref{Qhaar}))
is given by $Q_{R} = .988327$.  Therefore, when we compare the size of
the gap (hence the convergence rate) for the different topologies it
is important to remember that the closed chain is converging to a
slightly different asymptotic state than the other topologies.

Note that the same conservation of even or odd number of non-identity
Pauli operators occurs in the AA topology for an odd number of
qubits. This is because each qubit is involved in an even number of CZ
gates. In this case, however, initial computational basis states
evenly populate the two subspaces, therefore the final state is the
ergodic state.

%%%%%%%%%%%%%%%%%%%%%%%%%%%%%%%%%%%%%%%%%%%%%%%%%%%%%%%%%%%%%%%%%

\subsubsection{Numerical results}

\begin{figure}[t]
\includegraphics[width=8cm]{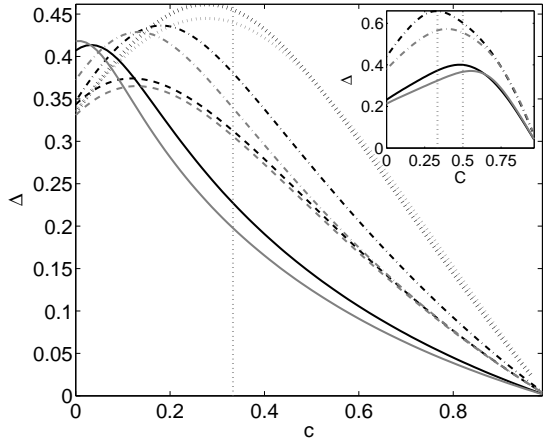}
\caption{Markov spectral gap, $\Delta$, versus local gate parameter,
$c$, for various qubit topologies using two-qubit CZ (main panel) and
XY gates (inset). Dark lines: $n=6$; Light lines: $n=8$.  For CZ
gates, the open chain topology (solid lines) has a larger gap than any
of the other topologies for single-qubit HZ gates, $c = 0$, but the
smallest gap for single-qubit random gates, $c = 1/3$ (vertical dotted
line). The HZ gate is the optimal single-qubit gate for the open chain
topology.  The AA topology (\dots) has the smallest or second smallest
gap (depending on the number of qubits) for HZ gates but the largest
for random gates. Random gates are close to optimal single-qubit
rotations for AA topology.  The closed-chain topology ($-\cdot$) has
the second largest gap for both random rotations and HZ gates. The
optimal single-qubit rotation for this topology would be $c \approx
.18$.  The star topology (dashed line) has the second smallest gap for
both random and HZ single-qubit rotations and the optimal single-qubit
gate (for 8 qubits) is also $c\approx .18$.  For random single-qubit
gates, going from 6 to 8 qubits decreases the size of the gap for all
topologies.  For single-qubit HZ gates, the gap increases when going
from 6 to 8 qubits for the open- and closed- chain topologies but
decreases for the star topology.  Inset: $\Delta$ versus $c$ for two
iterations (such that all couplings have been utilized) of PR
algorithms using two-qubit XY gates. The closed-chain topology
($-\cdot$) has a significantly larger gap than the open chain (solid
line). The optimal single-qubit gate for 8 qubits is the random
rotation, $c \approx 1/3$, for the closed chain, but $c \approx 1/2$
for the open chain.  Note the smaller gap (hence slower convergence
rate) of $n=8$ as compared to $n=6$.  }
\label{Mtop}
\end{figure}

Fig. \ref{Mtop} summarizes the behavior of the Markov spectral gap,
$\Delta$, for each topology as a function of $c$.  The optimal
single-qubit gate is dependent both on the topology and number of
qubits.  A certain single-qubit rotation may be exceedingly good for a
certain topology, but extremely poor for another topology.  Increasing
the number of qubits from $n=6$ to 8 affects the convergence of
different topologies in different ways: For certain topologies, this
increase in the number of qubits causes the gap to decrease, while for
other topologies the gap increases for certain single-qubit rotations
but decreases for other single-qubit rotations.

We first address the effect of topology on PR algorithms using random
single-qubit rotations, $c=1/3$.  The Markov analysis for 8 qubits
suggests that the AA topology should converge the fastest, followed by
the closed-chain topology (which, as mentioned above, converges to a
different state).  The fast convergence of the AA topology is not
surprising given that this topology allows for simultaneous
interaction between all qubit pairs.  The open chain has the slowest
convergence rate. The star topology is slower than the closed chain,
despite the fact that each qubit is only two couplings away from every
other qubit.  The need to traverse the central qubit is likely to be
the bottleneck to the spread of entanglement. This conjecture is
supported by our later explorations of probabilistic application of
two-qubit gates.  For all topologies, increasing $n$ from $6$ to $8$
{decreases} the gap -- meaning that it takes longer for the algorithm
to converge to random states.

\begin{figure}[t]
\includegraphics[width=8cm]{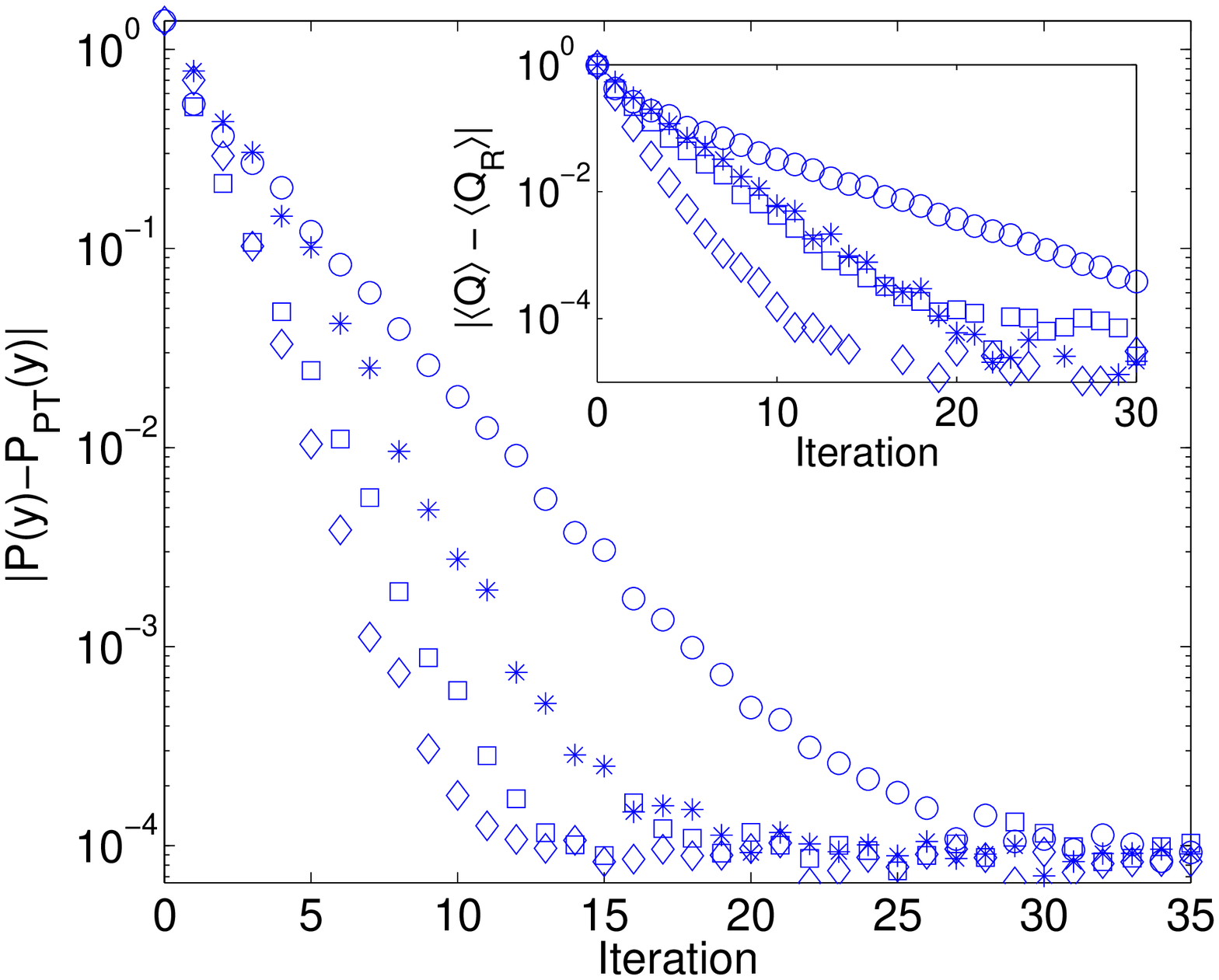}
\caption{(Color online) $l_2$-distance between PR distribution of
squared moduli of components in the computational basis and the
Porter-Thomas distribution (Inset: Distance of average global
entanglement from random value) for different qubit topologies using
CZ two-qubit gates and random single-qubit rotations (8 qubits, 100
implementations, all computational basis states). The qubits have been
arranged in an open chain ($\bigcirc$), a star formation ($*$), a
closed chain ($\square$), and in such a way that all the qubits are
connected to each other ($\diamondsuit$).  The convergence rate as a
function of topology agrees with Markov analysis, see
Fig. \protect\ref{Mtop}.  }
\label{topRandom}
\end{figure}

If random single-qubit gates are replaced by HZ gates, $c = 0$, the
open-chain topology exhibits the largest gap, followed by the closed
chain. In both cases, the convergence rate is faster than that of
random single-qubit gates, and {grows} when going from 6 to 8
qubits. The HZ gates also provide better PR algorithm convergence for
the star topology, though the rate is smaller for $n=8$ than $n=6$ in
this case.  Such a decrease in convergence when going from 6 to 8
qubits is more pronounced for single-qubit HZ gates than for random
ones.  Among the topologies we have examined, the AA topology is the
only one which demonstrates better behavior for random single-qubit
rotations than for the HZ gates.

\begin{figure}[t]
\includegraphics[width=8cm]{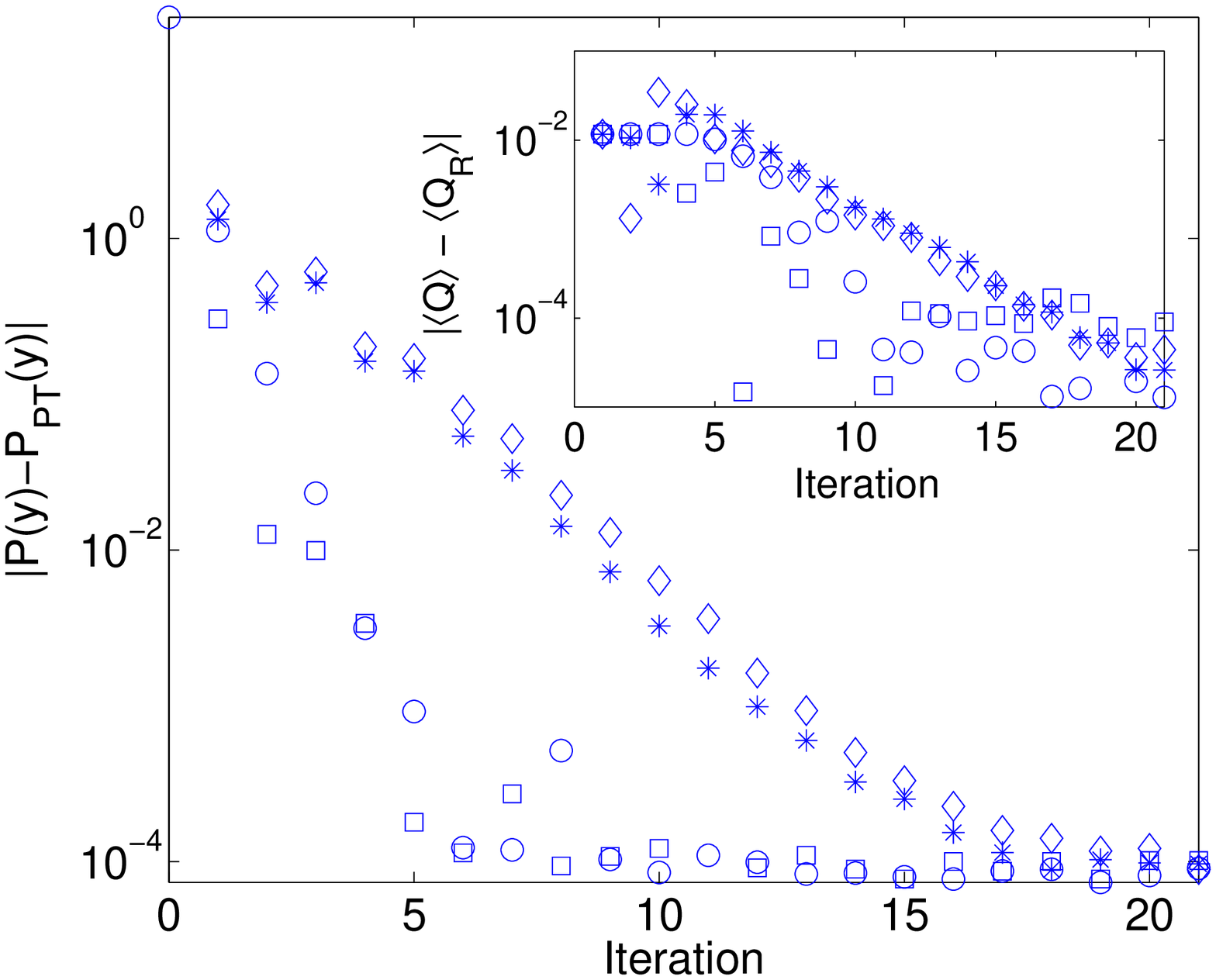}
\caption{(Color online) $l_2$-distance between PR distribution of
squared moduli of components in the computational basis and the
Porter-Thomas distribution (Inset: Distance of average global
entanglement from random value) for different gate topologies using CZ
gates and single-qubit HZ gates (8 qubits, 100 implementations, all
computational basis states).  The qubits have been arranged in an open
chain ($\bigcirc$), closed chain ($\square$), a star formation ($*$),
and in such a way that all the qubits are connected to each other
($\diamondsuit$). The decay rates of the star and AA topologies are
now slower than that of the open and closed chain, as expected from
the Markov analysis, see Fig.~\protect\ref{Mtop}.
%The open and closed chain have similar decay
%rates despite the predicted larger gap for the open chain.
%Also note the cut-off phenomenon described in \protect\cite{BWV}
%in the entanglement convergence rate for the open and closed chain
%topologies.
}
\label{topHz}
\end{figure}

The convergence to random state-vector element distribution and
entanglement for PR circuits using random and HZ single-qubit gates is
illustrated in Fig. \ref{topRandom} and Fig. \ref{topHz},
respectively.  For both quantities in both cases, the convergence
behavior is in agreement with the predictions based on Markov
analysis.  In particular, one sees that when using HZ gates, the AA
and star topologies become the slowest, and the open- and closed-
chain topologies have similar convergence rates. 
%despite the larger gap exhibited by the open-chain Markov analysis.  
Also notice that both
the open- and closed-chain topologies exhibit the entanglement `cutoff
phenomenon' described in \cite{BWV}. For the open (closed) chain,
every realization of the PR algorithm for $n/2$ ($n/2-1$) iterations
is maximally entangled. Only after this point does exponential decay
to random entanglement values set in \cite{remark3}.

\begin{figure}[t]
\includegraphics[width=8cm]{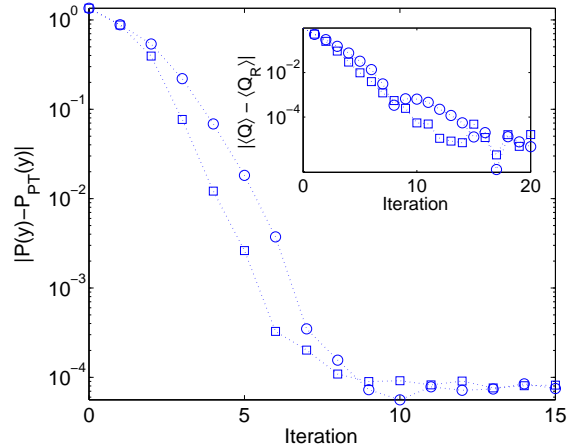}
\caption{(Color online) $l_2$-distance between PR distribution of
squared moduli of components in the computational basis and the
Porter-Thomas distribution (Inset: Distance of average global
entanglement from random value) for different gate topologies using XY
two-qubit gates and random single-qubit rotations (8 qubits, 100
implementations, all computational basis states). The qubits have been
arranged in an open chain ($\bigcirc$) and closed chain
($\square$). As expected, the convergence rate of the closed chain is
faster than that of the open chain.  }
\label{topXYRand}
\end{figure}

Results on the convergence of output state element distribution and
entanglement for PR circuits employing XY gates are given in
Fig. \ref{topXYRand}. As above, in this case the PR circuit implements
every other NN coupling for one iteration.  Similar to CZ gates, the
closed-chain topology does not result in convergence to the random
state. Nevertheless, the convergence rate of the closed chain to
random entanglement values and element distributions is faster than
that of the open chain.

In summary, the optimal single-qubit distribution for each of the
topologies exploiting two-qubit CZ gates is as follows: for the open
chain, the HZ gates are optimal, $c = 0$.  Remarkably, these gates are
naturally motivated by cluster-state QC \cite{BWV}.  For the AA
topology, the random single-qubit rotation is optimal, whereas for the
closed-chain and star topologies the optimal single-qubit rotations
correspond to $c \approx.18$.  For chain topologies, the optimal
$c$-value is also dependent on the number of qubits.  If two-qubit XY
gates are used, gate distributions with $c\approx 1/2$ and $c\approx
1/3$ are found to be optimal for open and closed chains, respectively.

\subsection{Probabilistic gates}

When PR algorithms employ two-qubit CZ gates, commutation allows any
number of such gates to be performed in parallel. Therefore, until
this point, we have constructed algorithms that maximize the number of
possible CZ gates based on the qubit topology per iteration. We
proceed to explore what happens if CZ gates are applied
probabilistically, that is, any specific CZ gate is applied with
probability $p$. As we shall see, for certain topologies lowering $p$
actually {\em increases} the convergence rate of the PR circuit.

\begin{figure}[t]
\includegraphics[width=8cm]{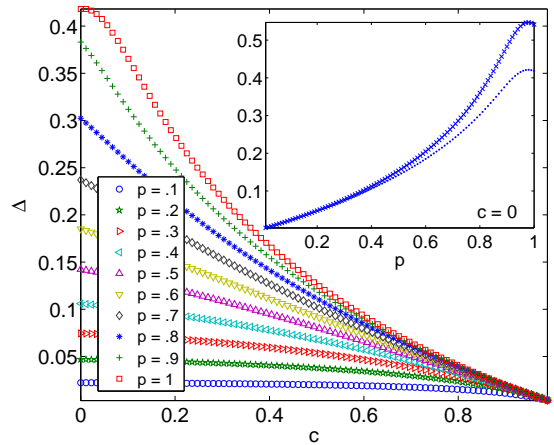}
\caption{(Color online) Markov spectral gap, $\Delta$, versus local
gate parameter, $c$, for an 8-qubit open-chain topology for different
probabilities, $p$, that each CZ gate will be implemented. The gap
increases with $p$ and the HZ gate ($c = 0$) is the optimal
single-qubit gate for all $p$ until $c = .98$.  Inset: Gap ($\cdot$)
and convergence rate ($\times$) for HZ gates as a function of $p$.
\label{OC8}}
\end{figure}

We begin by considering a probabilistic open chain topology.  As
illustrated in Fig.~\ref{OC8}, the Markov spectral gap (hence the
convergence rate) increases as $p$ is varied over $[0,1]$, $p=1$
recovering the deterministic case. The HZ single-qubit gate turns out
to be optimal for almost all $p$,
% For low values of $p$, the HZ gate is only
%slightly better than other single-qubit gates.
the improvement in the convergence rate being steady until $p$ is
close to unity.  We expect this since HZ gates are the `maximally
non-invariant' single-qubit gates with respect to the $z$-axis, which
is preferred by CZ gates.  As $p \rightarrow 1$, a rounded hump is
visible whose peak value for $n=8$ is found at $c = .02$ but whose
roundedness allows for a range of $c$ values giving nearly optimal
single-qubit gates. As the number of qubits increases, the peak of the
hump shifts towards $c = 0$, indicating that the non-optimality of the
HZ gates is due to the relatively small Hilbert space dimension. The
gap size and convergence rate of PR algorithms using HZ gates as a
function of $p$ are explicitly shown in the inset of Fig.~\ref{OC8}.

\begin{figure}[tb]
\includegraphics[width=8cm]{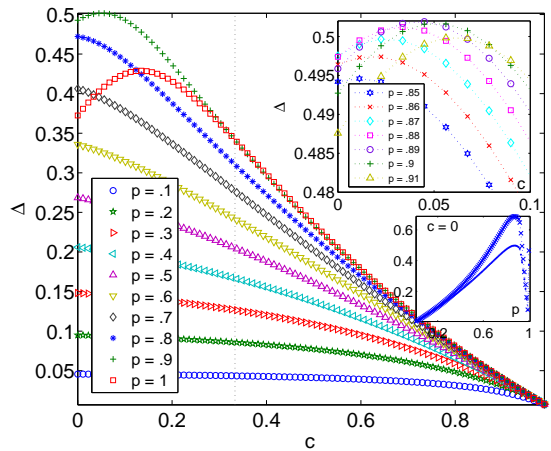}
\caption{(Color online) Markov spectral gap, $\Delta$, versus local
gate parameter, $c$, for an 8-qubit closed-chain topology with
probability $p$ that each CZ gate will be implemented. The HZ gate is
optimal until $p \simeq .85$. Above that, a hump grows towards
greater $c$ with increasing $p$. The height of the hump (size of the
gap) reaches a maximum at $p \simeq .89, c \simeq 4/90$ (see upper
inset). As $p$ continues to grow, the peak of the hump moves to higher
values of $c$ but now the size of the gap shrinks.  Lower inset: Gap
($\cdot$) and convergence rate ($\times$) for HZ gates as a function
of $p$. For $.9 < p < 1$, the decreased gap size is due to an
eigenvalue representing coupling between the two parity subspaces of
the system (see text) and thus is not reliable to fully determine the
actual convergence rate of the circuit.
\label{CC8}}
\end{figure}

The behavior of a probabilistic closed-chain topology is shown in
Fig.~\ref{CC8}.  Initially, as $p$ increases so does the gap and the
rate of convergence to random. The HZ single-qubit gate is optimal
until $p \simeq .85$ (for $n=8$).  As $p$ increases further, a hump
develops whose maximum increases until $p \simeq .89$, and decreases
afterwards.  The maximum of the hump shifts towards larger values of
$c$ as $p$ approaches one, resulting in optimized convergence at $p
\simeq .89, c \simeq 4/90$.  Observe that the optimal probability, $p
\simeq .89$, is close to $p = .875=7/8$: For an 8 qubit system, this
is the value for which, on average, there is one CZ gate per iteration
that is not implemented.  In other words, for this $p$ value the
closed chain is similar to an open chain with $p = 1$, except that the
opening, corresponding to the missed CZ gate, changes position
randomly at each iteration. This variable opening in the qubit chain
is what increases the convergence rate to the point where the maximum
attainable convergence rate on an open chain is outperformed by a
closed chain with $p > .7$. As the number of qubits increases, the
hump is pushed to higher values of $p$, consistent with the fact that
as $n$ grows, the chances of one two-qubit gate not being applied,
i.e. that there is a break in the chain, increases accordingly.

As noted above, the closed chain with $p = 1$ does not converge to
random due to the conservation of even or odd labeled non-identity
Pauli operators. However, for $p < 1$ this conservation rule does not
hold and ergodicity is restored.  Nevertheless, based on calculations
which directly compare the state of the system after many iterations
of the Markov matrix to the random state, it appears that a transition
occurs at $p \simeq .8$, roughly the value where the gap begins to
decrease. Below this point, the mixing between the two parity
subspaces is {\em on par} with the convergence to random within each
subspace. For $p \agt .8$, the relaxation between the subspaces
decreases -- eventually increasing the time for the PR circuit to
reach the final random state. This is demonstrated in the lower inset
of Fig.~\ref{CC8}, which shows the gap size and convergence rate of PR
algorithms using HZ single-qubit gates within the closed chain
topology as a function of $p$. For $.9 < p < 1$, the gap plummets and
then goes back up at $p = 1$. This small gap is due to the eigenvalue
corresponding to the rate of transfer between the two subspaces of the
system.  In terms of entanglement distribution, a fast convergence to
the value given by Eq.~(\ref{CCEnt}) occurs first, followed by a
slower convergence to the expected Haar value at a rate determined by
the small gap.

\begin{figure}[t]
\includegraphics[width=8cm]{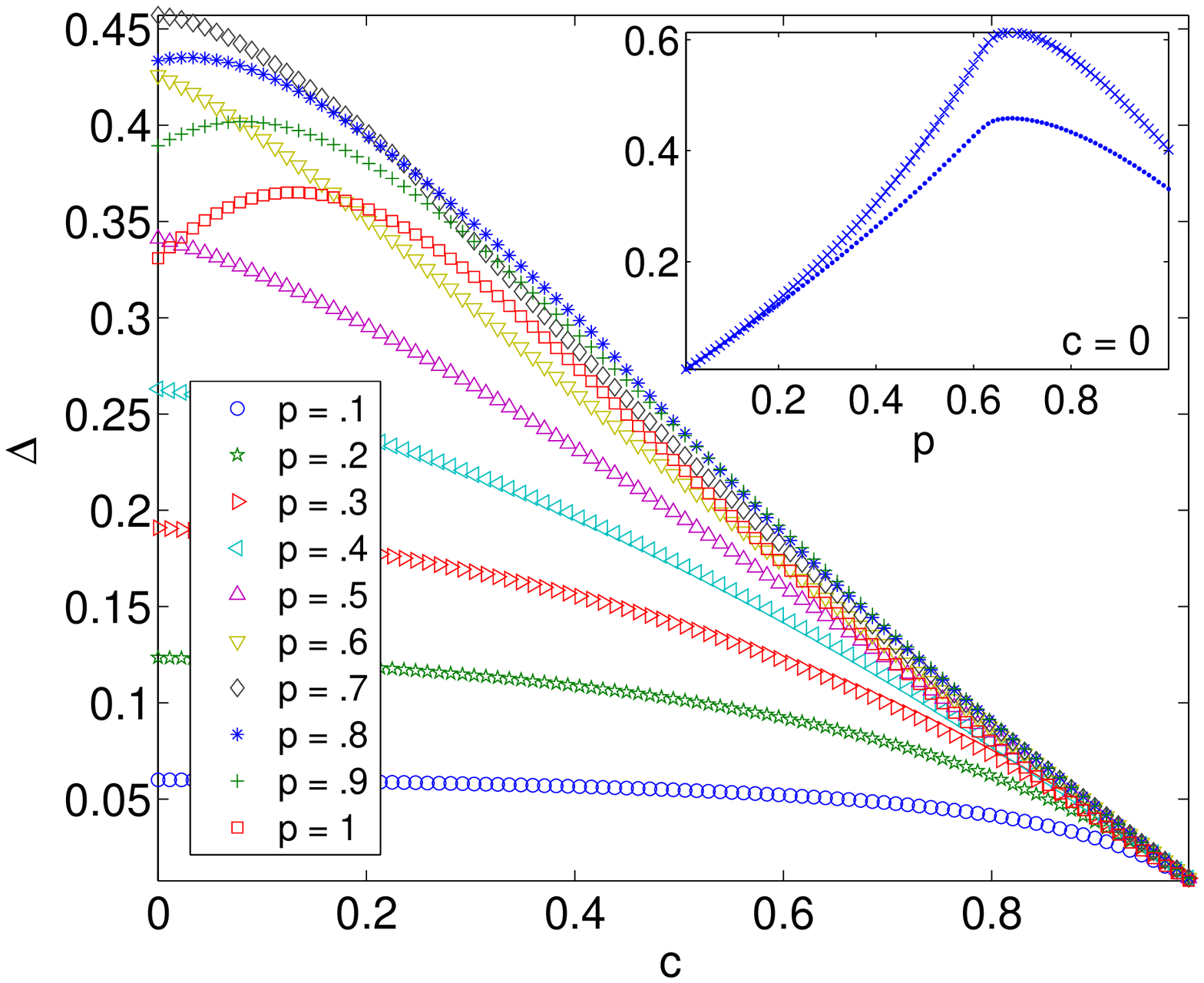}
\caption{(Color online) Markov spectral gap, $\Delta$, versus local
gate parameter, $c$, for an 8-qubit star topology with probability $p$
that a given CZ gate will be implemented. The gap increases with $p$
and the HZ gate ($c = 0$) is the optimal single-qubit gate for $p \alt
.7$. Inset: Gap ($\cdot$) and convergence rate ($\times$) for HZ gates
as a function of $p$.
\label{Star8}}
\end{figure}

Figure \ref{Star8} shows the Markov matrix gap for a probabilistic
star topology. For $p \alt .7$, the HZ is the optimal single qubit
gate, while for $p \agt .7$ a hump forms such that the maximum gap is
no longer at $c = 0$.  Unlike the closed-chain topology, the peak of
the hump is now always lower than the maximum overall gap, which is
found at $c = 0, p \simeq .7$ for $n=8$.  Note that for $.6 \alt p
\alt .8$, the convergence rate is faster than that of the open-chain
topology with $p = 1$.  The star topology has $(n-1)$ possible CZ
gates thus $p = .7$ means that there are about two inactive couplings
per iteration. The curves for the various values of $p$ appear to be
generally independent of the number of qubits.

\begin{figure}[t]
\includegraphics[width=8cm]{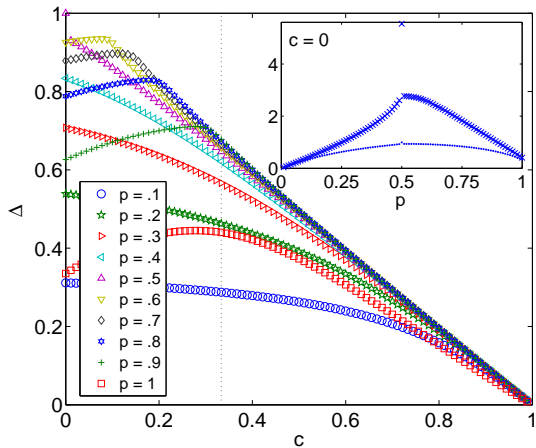}
\caption{(Color online) Markov spectral gap, $\Delta$, versus local
gate parameter, $c$, for an 8-qubit AA topology with probability $p$
that a given CZ gate will be implemented. The largest gap appears as a
singularity at $p = .5$ and $c = 0$. For $p < .5$, the gap increases
with decreasing $c$, the HZ gate ($c = 0$) being optimal. For $p >
.5$, the largest possible gap decreases and the optimal single-qubit
gate approaches random, $c = 1/3$, with increasing $p$. In addition,
the size of the gap as a function of $c$ exhibits a sharp change of
behavior after reaching its maximum. Inset: Gap ($\cdot$) and
convergence rate ($\times$) for HZ gates as a function of $p$. The
singularity at $p = .5$ is easily seen, as is the difference in the
behavior of the gap size as a function of $p$ above and below this
singularity.
\label{A2A8}}
\end{figure}

Lastly, a probabilistic AA topology is analyzed in Fig. \ref{A2A8}.
The observed gap values are much larger than any we have seen to this
point. The optimal single-qubit rotation once again depends on
$p$. For $p < .5$, the optimal single-qubit rotation is the HZ gate,
whereas for larger $p$ the optimal choice of $c$ increases until
reaching $c \simeq 1/3$ when $p = 1$. Unlike the other topologies, the
gap behavior as a function of $c$ for $p > .5$ exhibits a sharp change
after reaching a maximum. At $p = .5$ and $c = 0$, a sharp singularity
occurs in both $c$ and $p$, resulting in an extremely large gap,
$\Delta = .9961$. Remarkably, the size of this gap approaches one
exponentially as the number of qubits increases. We will explore this
specific case in more detail in the next section.

\section{Scaling behavior}

In this section we look at the effect of scaling on PR circuits with
certain qubit topologies. Specifically, symmetries in the AA and star
topologies allow for the computation of Markov matrix gaps for PR
circuits with a large number of qubits. In addition, utilizing the AA
topology with probabilistically applied two-qubit gates allows us to
give quantitative predictions on expected scaling behaviors for
different classes of qubit topologies based on their degree of
connectivity (as measured by how many qubits are on average connected
to each other via CZ gates).  This results in a simple formula for the
scaling of the spectral gap for each class, which is consistent with
the scaling results known to date.
%The complete permutation symmetry of the all to all topology makes it
%particularly efficient to simulate even for large numbers of qubits, allowing
%numerical results to be obtained regarding the asymptotic scaling of the gap
%with respect to the number of qubits.

\subsection{All-to-all topology}

Each time step of a PR circuit using an AA topology consists of
independent local gates on each qubit, followed by CZ gates between
arbitrary pairs of qubits.  We apply these gates probabilistically, in
such a way that for each possible pairing of qubits a CZ gate is
performed independently with probability $p$.

In order to access larger numbers of qubits, we construct a reduced
representation by considering equivalence classes of Pauli operators
under permutations of the qubits.  Thus, we label the states of the
reduced Markov matrix by the number of $Z$- and $X$-Pauli operators
they contain. As noted previously, although each eigenvalue of the
reduced chain occurs in the full representation, some eigenvalues of
the full chain do not occur in the reduced chain, allowing for the
possibility that the largest eigenvalue could, in principle, be removed.
In practice, we verified that for up to $n=10$
where comparisons with the full
representation are tractable, the eigenvalue that determines the gap
occurs in both representations.  Furthermore, the eigenvalue that
governs the decay of computational basis state must occur in the
reduced representation, because the second moments of computational
basis states are invariant under qubit permutations.

For fixed $p$, numerical results show that the scaling behavior of the
gap upon increasing the number of qubits differs depending on whether
$c=0$ or not.  For $c > 0$, the gap appears to converge at an
exponential rate to $1-c$,
\begin{equation}
\label{GapScale0}
\Delta^{(c >0)}(p ; n)=(1-c)- e^{-a(p)n},
\end{equation}
with a convergence rate $a(p)$ which is peaked at $p=0.5$. When $c=0$,
however, the gap scales as
\begin{equation}
\label{GapScale1}
\Delta^{(c=0)}(p \leq .5; n)=1-\frac{1}{2^{\alpha(p)p n+\beta(p)}},
\end{equation}
where $\alpha$ and $\beta$ are $p$-dependent factors, which are of order
unity for $p\le .5$. Thus, in this case the gap converges exponentially to one.

\begin{figure}[t]
\includegraphics[width=8cm]{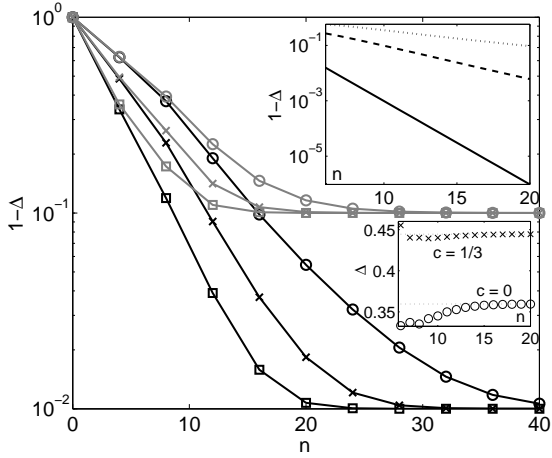}
\caption{$1-\Delta$, versus number of qubits for the AA topology using
probabilistically applied CZ gates. Dark curves: $c = .01$ and $p =
.25 (\bigcirc)$, .35 ($\times$), and .45 ($\square$). Light curves: $c
= .1$ using the same values of $p$. In both cases, the gap approaches
$1-c$ for increasing $n$, consistent with Eq.  (\ref{GapScale0}).
Upper inset: $1-\Delta$ versus number of qubits for the AA topology
when $c = 0$ and $p = .2$ (dotted line), .4 (dashed line), and .5
(solid line). The gap approaches one at an exponential rate for all
values of $p$ with the quickest convergence at $p = .5$, as captured
by Eqs. (\ref{GapScale1}) and (\ref{GapScale3}).  Lower inset: Gap
versus number of qubits for deterministic AA topology, $p = 1$, for
single-qubit HZ ($\bigcirc$) and random ($\times$) gates.  The gap
grows with increased $n$ until saturating at about $.3596$ for the HZ
gates and $.4444$ for the random gates.
\label{AAScaling}}
\end{figure}

Extremely interesting behavior of the gap occurs when $c = 0$ and $p =
.5$.  In this case, the eigenvalues of the Markov chain are found to
exhibit a very simple structure: there is the ergodic eigenvector with
eigenvalue 1, a single eigenvalue of $-2^{-n}$,
multiply degenerate eigenvalues of $2^{-n/2}$ and $-2^{-n/2}$, and all remaining
eigenvalues are zero.  Thus, the gap is given by
\begin{equation}
\label{GapScale2}
\Delta^{(c=0)}(p=.5;n)=1-\frac{1}{2^{n/2}}.
\end{equation}
This holds for {\em generic} initial states. Initial computational
basis states, however, have a non-zero component only along the
ergodic eigenvector (with eigenvalue 1) and the single eigenvector
with eigenvalue ${1}/{2^n}$.  This leads to a gap of
\begin{equation}
\label{GapScale3}
\Delta^{(c=0; \text{comp})}(p=.5;n)=1-\frac{1}{2^{n}},
\end{equation}
which is the fastest convergence rate for any PR circuit we have
examined.  The size of the gap as a function of the number of qubits
for $p \leq .5$ is shown in Fig.~\ref{AAScaling}.

We can use Eq.~(\ref{GapScale1}) to understand and generalize existing
results on the gap scaling behavior for PR circuits. First, let us
review what is known about the scaling behavior of the gap with regard
to the number of two-qubit gates applied per time step. For a PR
circuit in which a single two-qubit gate uniformly drawn from the Haar
distribution on SU(4) is implemented per time step between a pair of
qubits selected at random, an analytical result of
$$\Delta^{[1]}(n)= \Theta(1/n)$$
\noindent
was proved by Harrow and Low in \cite{Harrow08}.  In addition,
numerical work by Znidaric \cite{Znidaric} demonstrated the above
${1}/{n}$ scaling law to be {\em typical} for a large class of
two-qubit gates used in conjunction with random SU(2) single-qubit
gates.  For PR circuits where $\sim n$ commuting two-qubit gates are
performed in parallel per time step, numerical evidence suggests that
the gap tends to a constant value as the number of qubits increases
\cite{RM,BWV},
$$\Delta^{[n]}(n)=\text{const},$$
the value of such a constant depending in general on the type of gates and
qubit topology. For the AA topology under discussion here for fixed $p$ there
are $\sim n^2$ commuting two-qubit gates performed in parallel per time step.
We find for the case of $p<1$ and $c=0$ a scaling behavior
(as in Eq.~(\ref{GapScale1})) of the form
$$\Delta^{[n^2]}(n)=1-\frac{1}{2^{\kappa n + \gamma}},\;\;\; \kappa >0 ,$$
and constant scaling otherwise.

It is enlightening to compare the convergence rates for each of the
above cases in terms of the number of two-qubit gates applied rather
than the number of time steps.  Let $N_2$ denote the number of
two-qubit gates applied per time step.  In order to compare the gap,
$\Delta=1-\lambda_1$, for a protocol in which $N_2=1$ to one where
$N_2 \sim n$ (such as the open-chain, closed-chain, and star
topologies) or $ N_2 \sim n^2$ (such as in the AA topology), we
postulate an effective gap,
$$\Delta_{\text{eff}}^{[N_2]}=1-
\lambda_{\text{eff}}^{[N_2]},$$
\noindent
corresponding to a Markov matrix obtained by raising the original
Markov matrix for which there is only one two-qubit gate per time step,
to the power $ N_2$, so that the same
number of two-qubit gates are considered.
Presuming the gap for a protocol with $N_2=1$
is typically $\sim \frac{1}{n}$, it follows
that for $N_2\sim n$, in the limit where as $n$
is sufficiently large,
$$\lambda_{\text{eff}}^{[n]} = \lim_{n\rightarrow\infty}
\Big(1-\frac{\kappa}{n}\Big)^n=e^{-\kappa},$$
yielding a gap of
$\Delta_{\text{eff}}
^{[n]}=1-e^{-\kappa}$, which is a constant. For $N_2\sim n^2$,
this effective gap goes to $\Delta_{\text{eff}}^{[n^2]} =
1-e^{-\kappa n}$, which asymptotically approaches 1, consistent
with the above numerical results.

We may numerically explore all of the above cases within the AA
topology by allowing $p$, the probability of performing a two-qubit
gate between a given (arbitrary) pair of qubits, to be a function of
$n$. We consider the special case of $c=0$.  First, set
$$p^{[1]} = \frac{2}{n(n-1)},$$
so that on average one CZ gate
is performed per time step. We find numerically (for the AA
configuration) that the gap is given asymptotically by
$\Delta^{[1]}(n)=\frac{1.34}{n - 1.55}$, consistent with the above mentioned
results.

\begin{figure}[t]
\includegraphics[width=8cm]{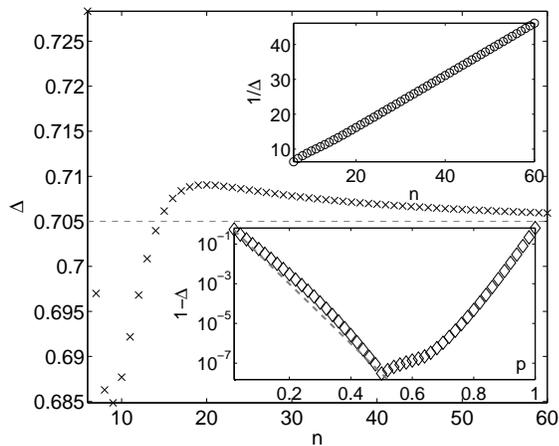}
\caption{Spectral gap, $\Delta$, of a PR circuit using an AA topology
as a function of the number of qubits, $n$, with two-qubit gates
applied with probability $p=\frac{2}{n-1}$ such that at each iteration
the number of gates performed is on average $n$. The gap
asymptotically approaches a constant value of $\approx .705$ (dashed
line). The upper inset shows the inverse gap, $1/\Delta$, as a
function of $n$ for $p=\frac{2}{n(n-1)}$ so that on average one
two-qubit gate is performed per iteration. The gap scales as $\propto
1/n$.  The lower inset shows $(1 - \Delta)$ for 50 qubits as a
function of $p$. The gap is largest, and hence convergence quickest,
at $p=.5$. The approximate formula for the gap when $p < .5$ is given
by $\Delta=1-1/2^{np}$ and is shown as a dashed line.}
\label{scale_n}
\end{figure}

Next, set
$$p^{[n]}=\frac{2}{n-1},$$ so that an average of $n$ CZ gates are
applied per time step. In the large-$n$ limit, this results
numerically in a constant gap of $\Delta^{[n]} = .705$. Finally, for
fixed $p \le 0.5$, the gap approaches one as in Eq.~(\ref{GapScale1}).
The above results are demonstrated in Fig.~\ref{scale_n}

In summary, we find that for a broad class of random circuits the
gap scales as $\Delta =
1-\exp(-\alpha N_2/n)$, $\alpha>0$.  Recalling that the convergence
rate, $\Gamma = -\ln(1-\Delta) = \alpha N_2/n$, we find that {\em the convergence rate scales
asymptotically as $\sim {1/n}$ per two-qubit gate.} This scaling
behavior appears to be typical for all PR circuits where $N_2=1$ to
$\sim n$ and for some types of PR circuits with when $N_2 \sim n^2$,
specifically when $p<1$ and $c=0$ for the AA topology. This is
important because if one wishes to parallelize by simultaneously
performing commuting two-qubit gates then it is possible to perform
the maximum $N_2 \sim n^2$ gates without sacrificing the typical $1/n$
scaling per two-qubit gate of the convergence rate.

\subsection{Star topology}

For qubits arranged in the star topology, CZ gates may be applied only
between the central qubit and the $(n-1)$ outer qubits. With the
proper choice of parameters, this structure also allows (as the AA
topology) for a simple specification of the Markov matrix eigenvalues.
Let single-qubit gates be drawn from a distribution with $c = 1/3$ and
each CZ gate be applied with a probability $p = 3/4$. Then, the
eigenvalues of the Markov matrix consist of a $(2^n - 2)$-fold
degenerate eigenvalue equal to $2/3$, and a pair of eigenvalues at
$\frac{1}{3}(1 \pm \frac{1}{2^{n-1}})$. The remaining eigenvalues are
all zero, leading to a gap of
$$\Delta^{(\text{c}_{\rm{all}}\rm{=1/3})}(p=.75, n) = 1/3.$$

We can also examine a situation where $p = 3/4$ but we draw the single
qubit rotation on the central qubit from a different distribution than
that of the other qubits.  If, for example, we draw the non-central
qubit rotations from a random distribution and the central qubit
rotation from a distribution parameterized by $c$, the corresponding
Markov matrix develops a $(2^n - 2)$-fold degenerate eigenvalue equal
to $\frac{1+c}{2}$, a pair of eigenvalues at $c\pm \frac{1-c}{2^n}$,
and the remaining eigenvalues are zero. The gap is now maximized at .5
when $c=0$ for the local gate distribution of the central qubit,
$$\Delta^{(\text{c}_{\rm{central}}\rm{=0},\rm{c}_{\rm{outer}}\rm{=1/3})}(p=.75,
n) = 1/2.$$ Interestingly, this PR circuit is very similar in
structure and performance to the approximate 2-design outlined in
\cite{EmersonDesign}, as discussed in Appendix B.

\section{Effect of collective rotations}

In all of the above analysis we have assumed PR circuits where the
single-qubit rotations applied to each qubit are drawn independently
from some specified distribution at each time interval. What would happen
if we limit the number of times we draw from the given distribution? For
example, will a PR circuit that applies the same random rotation to
every qubit (or every other qubit) during a given iteration converge
more slowly than if every qubit rotation is independent?

Constructing a PR circuit where we decrease the number of
independently applied
single-qubit random rotations may be viewed as an exercise in
generating random states (or unitary operators) with as few degrees of
freedom as possible.  It is well known that quantum chaotic systems
with only one or two degrees of freedom can mimic several properties of
random systems \cite{Haake}.  Exploring PR circuits by
reducing the number of degrees of freedom goes one step further, in
that it addresses a similar question within a general mathematical
framework rather then looking
at specific examples of what may be extreme cases. This aspect of
PR circuits has been previously explored in
\cite{YSWsymm} with respect to the unitary operators applied by the
algorithm. Here, we look at some further statistics of the generated
PR states not explored in the previous work and compare
them to the same characteristics of random states described above.

\begin{figure}
\includegraphics[width=8cm]{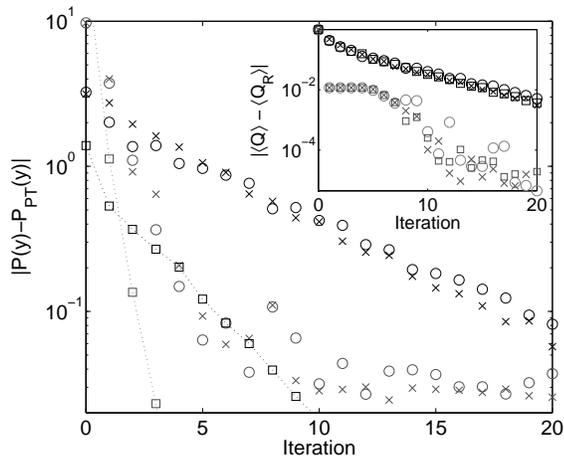}
\caption{\label{QCA} $l_2$-distance between PR distribution of squared
moduli of components in the computational basis and the Porter-Thomas
distribution (inset: Distance of average global entanglement from
random value) versus number of iterations for open-chain topologies (8
qubits, 100 implementations) in which each qubit undergoes equivalent
random rotations ($\bigcirc$) or every other qubit undergoes
equivalent random rotations ($\times$). Both random single-qubit gates
(dark) and HZ gates (light) were used. These are compared to circuits
in which each qubit undergoes a different rotation ($\square$).
Interestingly, using different rotations on all odd and all even
qubits does not improve the convergence rate beyond the one achievable
by using only one collective rotation for all qubits.  }
\end{figure}

Specifically, let us consider an open chain of qubits with CZ gates
between NN qubits. For the single-qubit gates, instead of applying
different random or HZ rotations to every qubit at every PR iteration,
we apply the same {\em collective} random or HZ rotation to all qubits
for a given iteration. We also look at a case where, instead of one
single-qubit rotation per iteration, we apply two different rotations,
the first to odd qubits and the second to even qubits.  We compare the
convergence of the element distribution and entanglement for these two
cases to that of an open chain where different rotations (random or
HZ, respectively) are applied to each qubit. As shown in
Fig. \ref{QCA}, limiting the diversity of single-qubit rotations
appears to have little effect on the convergence rate of
entanglement. The matrix element distribution, however, is strongly
affected by the symmetries, and the convergence to the PT distribution
is significantly slowed down by reducing the number of free parameters
in the PR algorithm.  Were a quantum protocol dependent on randomness in the state element distribution, these PR circuits would require significantly more steps than PR circuits with different random
rotations on each qubit.  It is worth noting that for the case in
which every qubit undergoes the same local gate, the PR gate set is
not universal over the full Hilbert space because it preserves parity
with respect to reversing the order of the qubits along the
chain. Nevertheless, similar to previous cases, the asymptotic decay
rate is not heavily influenced by this fact, although the final
asymptotic values of the test functions under consideration may differ
slightly from those expected for the Haar distribution.

\section{Initial State Dependence}

One of the challenges in constructing PR states is creating sufficient
entanglement using simple one- and two-qubit gates. If, however, the
initial state already has some entanglement the absolute convergence
(although not the rate of convergence) to randomness may proceed more
quickly.  To demonstrate this, we apply PR circuits to initial states
of the form:
\begin{equation}
\label{states}
|\psi(a)\rangle = \frac{1}{K(a)}\Big[(1-a)(|0\rangle^{\otimes
  n}+|1\rangle^{\otimes n})+ a(|0\rangle+|1\rangle)^{\otimes n}
\Big],
\end{equation}
where $K(a)$ is the normalization factor.  When $a = 0$, the state
$|\psi(0)\rangle$ is a generalized GHZ state which is maximally
entangled relative to arbitrary local observables, and $K(a) =
1/\sqrt{2}$. When $a = 1$, $|\psi(1)\rangle$ is completely separable,
and $K(a) = 1/\sqrt{N}$.  By applying the PR circuit to initial states
with different values of $a$, we can explore the number of iterations
gained by using an initial entangled state. Of course, starting with
an initially entangled state has a cost. The cost is minimal, however,
since the above states are easily generated via a series of controlled
$\sigma_x$-rotations by angles which are dependent on $a$. These gates
can be implemented simultaneously since all gates share the same
control qubit.

\begin{figure}
\includegraphics[width=8cm]{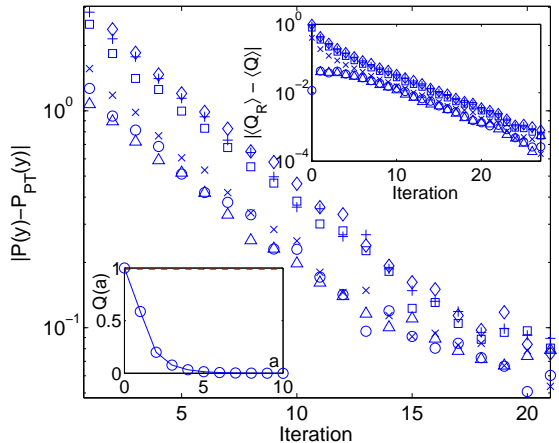}
\caption{\label{EI} (Color online) $l_2$-distance between PR
distribution of squared moduli of components in the computational
basis and the Porter-Thomas distribution (inset: Distance of average
global entanglement from random value) versus number of iterations of
a PR algorithm for 8 qubits (500 samples) and initial states $a = 0$
($\bigcirc$), $a \simeq .02337$, corresponding to a state with random
entanglement ($\triangle$), .1 ($\times$), .2 ($\square$), .3 ($+$),
and 1 ($\diamondsuit$). All states converge to random at the same
rate. Note that the $a = 0$ case has a 4 or 5 iteration advantage over
the $a = 1$ initial state.  Lower inset: Entanglement of the initial
states versus $a$.  The random entanglement value is $\simeq .9883$.
}
\end{figure}

Figure \ref{EI} shows the difference in state element distribution and
average entanglement between random states and states produced from
the PR circuit with different $a$ initial states as a function of
iteration.  We have used a PR algorithm with an open-chain topology,
CZ gates, and random single-qubit rotations. All states approach the
random state entanglement at the same exponential rate. Not too
surprisingly, however, initial states with entanglement closer to that
of random states approach random state entanglement faster than other
states. For $n = 8$, the entanglement of the state $a \simeq .02337$
equals that of random states, and has the fastest convergence to
random entanglement (that the state with entanglement equal to that of
random states converges the fastest is more clearly seen for smaller
numbers of qubits). As the number of qubits grows, the random
entanglement gets closer to maximum, therefore the GHZ state provides
optimal convergence.  Perhaps more surprisingly, the states with more
initial entanglement also converge to the random state element
distribution more quickly than their less entangled counterparts.
This is in-line with the notion that, in general, characteristic
entanglement and state element distribution are correlated
\cite{YSW3}.

Figure \ref{EI} allows us to quantify the advantage of starting with
entangled states in terms of the number of iterations by noting where
the different curves cross a given distance from random entanglement
level. For example, the $a = 0$ state gives an advantage of 4-5
iterations over the $a = 1$ state.

To further test the dependence of a PR algorithm convergence on the
initial state, we have also used as the initial state the open-chain
cluster state (still utilizing a circuit model PR algorithm). We find
that the results are equivalent to that of the generalized GHZ
state. Note that with respect to the multi-partite entanglement
measure we are using, both these states are maximally entangled.
However, the apparent equivalence between the GHZ state and cluster
state does {\em not} hold for all topologies. For example, the
convergence of a PR algorithm with the qubits in the AA topology does
not benefit from an initial GHZ state, whereas it does benefit from an
initial cluster state.

\section{Conclusion}

We have studied the efficiency of PR circuits as a function of
single-qubit and two-qubit gates, qubit topology, the probability of
applying the two-qubit gates, and system size, as well as addressed
the effect of collective rotations and different initialization. Using
a Markov chain analysis, we have analyzed the strong interdependence
of the choice of single- and two-qubit gates. The optimal PR algorithm
will consist of a distinct pair of single- and two-qubit gates. The
optimal choice of gates is also dependent on the qubit
topology. Different topologies will favor one or another single- or
two-qubits gates.  Also dependent on the topology, the efficiency of
the PR algorithm may increase if the two-qubit gates are applied
probabilistically.  In particular, we noticed extremely fast
convergence for the AA topology with $p = .5$.

\iffalse
We have also studied PR circuits in which we apply
collective rotations as opposed to individual
rotations on each qubit, and noted the affect on the PR circuit
of using entangled initial states.
\fi

As one of the main implications of our work, we have demonstrated how,
as the number of qubits is increased, the asymptotic convergence rate 
of second order moments of the PR circuit scales as $1/n$ per two-qubit gate 
implemented per time step.  This may prove useful
to quantitatively characterize PR circuits in terms of $t$-designs. In
general, obtaining a deeper understanding of the behavior of PR
algorithms as approximate higher-order $t$-designs appears to be an
important next step toward harnessing quantum pseudo-randomness.

\acknowledgments

It is a pleasure to thank David G. Cory, C. Stephen Hellberg, and
Cecilia C. Lopez for valuable discussions and input. WGB gratefully
acknowledges partial support from Constance and Walter Burke through
their Special Projects Fund in QIS.  YSW acknowledges support from the
MITRE Technology Program under MTP grant \#07MSR205.

\appendix

\section{Cluster-state parameters}

Several potential applications for PR circuits are in the area of
quantum communications, which is naturally suited for photonic
implementations. Current analyses suggest that linear optics
QC will require too many resources, in terms of amount of
optical hardware, for practical implementations \cite{N}. The
cluster-state approach to QC, however, may be less stringent on the
number of necessary elements.

A cluster state \cite{BR3} can be created by first rotating all
qubits into the state $\frac{1}{\sqrt{2}}(|0\rangle +
|1\rangle)$. Qubits are then entangled by applying CZ gates between
pairs of qubits $j$ and $k$.  In a graphical picture of a cluster
state, qubits are represented by circles and pairs of qubits that have
been entangled via a CZ gate are connected by a line.  A cluster state
with qubits arranged in a two-dimensional lattice such that each qubit
has been entangled with four NNs, suffices for universal
QC \cite{BR1}. For photonic realizations of cluster-state QC, where
two qubit gates can only be applied probabilistically, one assumes an
unlimited number of maximally entangled EPR pairs (attained for
example via spontaneous parametric down-conversion) which can be used
to construct larger cluster states using so-called fusion operations
\cite{BR}.

Once the desired cluster state has been constructed, any QC algorithm
can be implemented using only single-qubit measurements in the $x$-$y$
plane. The inherent entanglement in the cluster state means that
measurement on one qubit may effect the state of the remaining
qubits. Thus, one can view each row of the cluster-state lattice as
the evolution of a single-qubit in time. Connections between rows are
used for two-qubit gates. Processing measurements are performed by
column from left to right until the last column (left unmeasured),
which contains the output state of the quantum algorithm, to be
extracted by a last readout measurement.  For a one-dimensional
cluster chain, the logical operation implemented by measurement along
an angle $\phi$ in the $x$-$y$ plane is $X(\pi m)HZ(\phi)$, where $H$
is the Hadamard gate and $Z(\alpha)$ ($X(\alpha$)) is a $z$- ($x$-)
rotation by an angle $\alpha$ \cite{BR2}. The dependence of the
logical operation on the outcome of the measurement is manifest in $m
= 0, 1$ for measurement outcome $-1, +1$. Any arbitrary rotation can
be implemented via three logical single-qubit rotations of the above
sort yielding
\begin{equation}
HZ(\alpha+\pi m_{\alpha})X(\beta + \pi m_{\beta})
Z(\gamma + \pi m_{\gamma}),
\end{equation}
where $(\alpha, \beta, \gamma)$ are the Euler angles of the
rotation. By drawing the Euler angles according to the Haar
measure, a random single-qubit rotation can be
implemented. Because the Haar distribution on SU$(2)$ is invariant
under $\pi$-rotations, the measurement-dependent
$\pi$-rotations may be ignored.
Two-qubit gates are
performed via a connection between two rows of the
cluster state. CZ gates in particular are `built-in' to the cluster
state and simple measurement automatically implements the gate.

PR state generation via the cluster state model of QC was first
explored in \cite{BWV}. The algorithm was initially performed using
three qubits in a chain to implement random rotations, $c = 1/3$, and
a connection between rows of the cluster state on every third qubit
implemented the CZ gates. In this way, $\ell$ iterations of an
$n$-qubit PR circuit required a lattice of $n\times 3\ell+1$ qubits
(where the extra $1$ comes from the final, unmeasured
column). However, by `filling in' the extra vertical connections, thus
making the single-qubit rotations HZ gates with $c = 0$, such that all
qubits were connected to their four NNs, the convergence rate of the
PR algorithm increased.

\subsection{Cluster topologies}

We explore cluster state topologies where the `columns' of the
two-dimensional lattice are connected as circles (making the cluster
state a cylinder), where all qubits in a `column' are connected to
each other (AA) and where each qubit is attached only to one central
qubit (stars). We note that both the AA shape and the star shape are
graph representations of generalized GHZ states differing only by
single-qubit rotations \cite{HEB}. Within the star topology, we look
at two different measurement strategies -- measuring the central qubit
first, and measuring the central qubit last.

\begin{figure}[t]
\includegraphics[width=8cm]{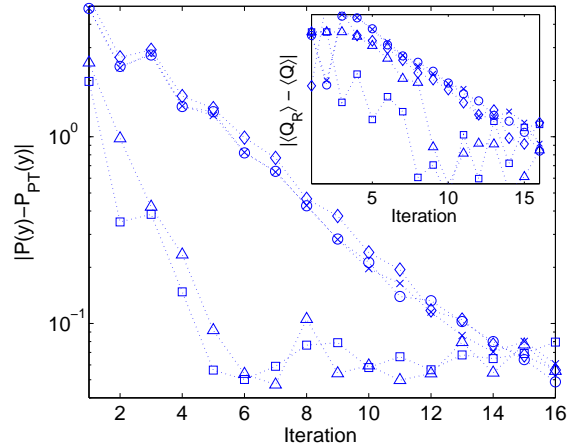}
\caption{(Color online) $l_2$-distance between PR distribution of
squared moduli of components in the computational basis and the
Porter-Thomas distribution (inset: Distance of average global
entanglement from random value) versus number of iterations for
different cluster state topologies (8 qubits, 500 implementations):
2-dimensional lattice ($\triangle$), cylinder ($\square$), star
topology measuring the central qubit first ($\times$), star topology
measuring the central qubit last ($\bigcirc$), and AA
($\diamondsuit$). The cylindrical lattice converges most quickly.  The
GHZ-type topologies (star and AA) converge more slowly and are, in
this way, like the AA topologies of circuit-model PR algorithms using
two-qubit CZ and one-qubit HZ gates. The open and closed chain exhibit
cutoff behavior in the entanglement convergence, as noted in
\protect\cite{BWV}.
\label{clustertop}}
\end{figure}

Figure \ref{clustertop} shows that the GHZ-type states converge more
slowly than the two-dimensional lattice and cylindrical topologies
with the cylindrical topology converging fastest. This is similar to
the circuit model for PR algorithms employing two qubit CZ gates and
one qubit HZ gates where the AA topology converges the
slowest.
%The difference is that in the cluster algorithm the
%cylindrical topology converges faster than the two-dimensional lattice
%whereas in the circuit model the open chain converges more quickly
%than the closed chain topology.
%%WB: There is not really a difference between the Cluster-state algorithm
%and the corresponding circuit algorithms, (they may be mapped onto each other),
%thus the last statement can't be quite right.  The fact that the gap for the
% open chain when c=0 is larger than that for the closed chain means that
%asymptotically the converegcne rate for Q must be faster for the 2D lattice
%than for the cylindrical cluster state. Since Markov does not completely
%specify P(y)(It can only describe its second moment) it is possible that the
%convergence rate of P(y) is larger for the cylinder than the 2D lattice
%but that must also be true for the converegne rate of P(y) for the closed vs open chains with c=0.

\subsection{Probabilistically applied cluster gates}

Photonic implementations of cluster state QC require construction of
the appropriate cluster state using operations that work
probabilistically. Thus, any cluster state has an associated cost in
the amount of resources required to, on average, build the given
state. Much work has gone into finding algorithms that will lower the
cost of construction for cluster states of two-dimensions, i.e.~in
which each qubit has 4 NNs \cite{ClusterConst}. In these works, cost
is generally associated with the number of EPR pairs needed for state
construction.  One conclusion that emerges from this work is that the
cost of constructing one dimensional chains is relatively low and it
is joining these chains (such that CZ gates can be applied) which is
the majority of the cost. For this analysis we assume that these
connections can be attempted without breaking the chain (even if the
operation is not successful).

We have already noted that a full two-dimensional cluster state (with
every vertical connection filled in) is optimal for PR
cluster-circuits. However, what if we were to apply the cluster CZ
gates only probabilistically? While we know from our exploration of
circuit model PR algorithms using an open chain topology that the
number of necessary iterations would increase, a probabilistic
application of these gates would cut the construction cost of the
initial cluster state.

Let us assume that an attempt to link two cluster chains together
succeeds with probability $p'$. We also assume the availability of
infinitely long cluster chains. As an example let $p' =1/2$ and we
attempt to construct a cluster state which can implement $C$
iterations of the cluster PR circuit in which the two qubit gates are
implemented with probability $p$. On average, construction of such a
cluster state would require $2Cpn$ probabilistic operations, where $n$
is the number of rows (logical qubits) in the cluster lattice. Thus,
for an 8-qubit cluster PR algorithm, to construct a cluster capable of
implementing a PR circuit with a convergence rate of .547 (which is at
$p = .98$ as shown in Fig.~\ref{OC8}) requires on average $1.96Cn$
probabilistic operations (recall that the circuit PR algorithm with HZ
gates is equivalent to the cluster-state PR algorithm).  A PR circuit
with half that convergence rate, would require $2C$ iterations but
could use an algorithm with $p \simeq .705$. This would require on
average $\simeq 2.82Cn$ attempts. Thus, for this case, it is better to
construct a cluster state capable of higher $p$ value PR
circuits. However, further analysis is necessary in the more realistic
case where all resources are accurately counted (including losses
taken upon failure of fusion operations), for different topologies,
and for other values of $p$ and $p'$.

\section{Twirling as a pseudo-random circuit}

Clifford twirls \cite{DLT}, or 2-designs, play an important role in a
number of quantum information protocols. A construction for an
approximate Clifford twirl has recently been suggested in
\cite{EmersonDesign}.  The procedure detailed in \cite{EmersonDesign}
consists of a sequence of twirling operations. A twirl consists of i)
performing one operation out of a specified set of operations; ii)
allowing the super-operator, $\Lambda$, to occur; iii) undoing the
original operation. The net effect of the twirl is to convert
$\Lambda$ into a new super-operator $\Lambda'$.  In the case of a
Clifford twirl, in which the set of operations is the Clifford group,
any super-operator is converted into a depolarizing channel.

The effect of the sequence of twirling operators on the set of Pauli
strings can be represented as a Markov matrix, identical to those that
describe the decay of the second moments of a PR circuit. Thus, a
one-to one mapping exists between PR circuit constructions and
approximate twirling operations. When the procedure in
\cite{EmersonDesign} is considered as a PR circuit, it bears a close
resemblance to PR circuits in a star topology, with probabilistic CZ
gates, and random local gates applied to the non-central qubits.
However, the procedure of \cite{EmersonDesign} involves three time
steps each of which is counted as an application of simultaneous
probabilistic CZ gates.  An approximate calculation of the convergence
rate is made in \cite{EmersonDesign} in order to determine scaling
properties. We find that an exact calculation of the gap of the
corresponding Markov matrix gives $\Delta_{\text{Clifford}}(c=1/3) =
5/6$, irrespective of the number of qubits.  We can improve the
convergence rate of this procedure marginally by using the following,
simpler construction: perform a PR circuit with the star topology with
$c=0$ for the central qubit, while keeping $c=1/3$ for the non-central
qubits and a gate probability of $p=3/4$. We find a gap of
$\Delta=1/2$.  Since this must be applied three times to perform the
same number of gates as in \cite{EmersonDesign}, this improved
construction yields an effective gap of $\Delta_{\text{Clifford}}(c=0)
= 1-1/2^3= 7/8$.


\begin{thebibliography}{99}
\bibitem{RemarkXY} If no restriction is placed on allowable pairs, CZ
and XY gates yield an equivalent convergence rate when used in
conjuction with SU(2) single-qubit gates.
\bibitem{remark3} The data is noisy due to the two-fold degeneracy of
$\lambda_2$, the largest non-trivial eigenvalue of $M'$.  When
smoothed, the decay is exponential as expected.
\end{thebibliography}
\end{document}